\documentclass[a4paper,11pt]{article}
\pdfoutput=1 

\usepackage{jheppub} 
\usepackage{etoolbox}                     
\makeatletter
\patchcmd{\maketitle}{\@fpheader}{Prepared for submission to JHEP}{}{}
\makeatother

\usepackage[T1]{fontenc}
\usepackage{amsmath,amssymb,mathtools,slashed}
\usepackage{color,xcolor}
\definecolor{blue}{rgb}{0,0,0.5}
\usepackage[utf8]{inputenc}
\usepackage{enumerate}
\usepackage[printonlyused]{acronym}
\usepackage{footnote}
\makesavenoteenv{tabular} 
\makesavenoteenv{table} 
\usepackage{hhline}

\usepackage{multirow}

 \definecolor{darkgreen}{rgb}{0.1,0.1,1.0}
 \hypersetup{
    linktocpage,
     colorlinks,
     citecolor=darkgreen,
     linkcolor= darkgreen,
     urlcolor=darkgreen
 }

\renewcommand{\arraystretch}{1.5}
\renewcommand{\Im}{\textrm{Im}}

\newcommand{\al}{\alpha}

\newcommand{\ga}{\gamma}
\newcommand{\Ga}{\Gamma}

\newcommand{\eps}{\epsilon}

\newcommand{\la}{\lambda}
\newcommand{\La}{\Lambda}

\newcommand{\Om}{\Omega}
\newcommand{\sig}{\sigma}

\newcommand{\als}{\al_s}
\newcommand{\NC}{N_C}

\newcommand{\mukin}{\mu_\pi^2}
\newcommand{\muG}{\mu_G^2}
\newcommand{\darwin}{\rho_D^3}

\newcommand{\Dleft}{\overset{\leftarrow}{D}}

\newcommand{\diquark}{\mathcal{D}}

\newcommand{\intm}{\textrm{int}^{-}}
\newcommand{\intp}{\textrm{int}^{+}}
\newcommand{\exc}{\textrm{exc}}

\newcommand{\Bary}{\mathcal{B}}
\newcommand{\Barycc}{{\mathcal{B}_{cc}}}
\newcommand{\Baryccstar}{{\mathcal{B}^*_{cc}}}

\newcommand{\Mes}{M}

\newcommand{\Omc}{\Om_c^0}
\newcommand{\Omcc}{\Om_{cc}^{+}}
\newcommand{\Omccstar}{\Om_{cc}^{+,*}}
\newcommand{\Xiccpp}{\Xi_{cc}^{++}}
\newcommand{\Xiccppstar}{\Xi_{cc}^{++,*}}
\newcommand{\Xiccp}{\Xi_{cc}^{+}}
\newcommand{\Xiccpstar}{\Xi_{cc}^{+,*}}
\newcommand{\Xicc}{\Xi_{cc}}
\newcommand{\Xiccstar}{\Xi_{cc}^{*}}

\newcommand{\Opsix}[2]{O_{#1}^{#2}}
\newcommand{\Opsixt}[2]{\tilde{O}_{#1}^{#2}}

\newcommand{\OpsevenP}[2]{P_{#1}^{#2}}
\newcommand{\OpsevenPt}[2]{\tilde{P}_{#1}^{#2}}

\newcommand{\mQheavy}{m_Q}
\newcommand{\Lifetime}[1]{\tau \left( #1 \right)}

\newcommand{\Psipc}{\Psi_{c}^{\dagger}}
\newcommand{\Psic}{\Psi_{c}}

\newcommand{\fs}{\,\textrm{fs}}

\newcommand{\GeV}{\,\textrm{GeV}}
\newcommand{\MeV}{\,\textrm{MeV}}

\definecolor{ivn}{rgb}{0.1,0.6,0.1}
\definecolor{JG}{rgb}{0.1,0.2,0.9}

\title{Revisiting lifetimes of doubly charmed baryons}
 \author[a]{Lovro Dulibi\' c,}
 \author[a]{James Gratrex,}
 \author[a]{Bla\v zenka Meli\' c,}
 \author[a]{and Ivan  Ni\v sand\v zi\'c}

\affiliation[a]{Ru\dj er Bo\v skovi\'c Institute, Bijeni\v cka cesta 54, 10000, Zagreb, Croatia.}

\emailAdd{ldulibic@irb.hr}
\emailAdd{jgratrex@irb.hr}
\emailAdd{melic@irb.hr}
\emailAdd{ivan.nisandzic@irb.hr}

\abstract{
We present updated predictions for lifetimes of doubly charmed baryons, within the heavy quark expansion, including available NLO $\alpha_s$ contributions and newly-computed terms in the $1/m_c$ series.
Our improved results confirm the expected hierarchy $$\tau(\Xi_{cc}^{+})  < \tau(\Omega_{cc}^{+}) < \tau(\Xi_{cc}^{++}) \,, $$ while the predicted lifetime 
 $\tau(\Xi_{cc}^{++}) = 0.32 \pm 0.05 ^{+0.08}_{-0.07} \,\textrm{ps} $ is consistent with the recent LHCb determination. We provide predictions for the lifetime ratios of the $\Xiccp$ and $\Omcc$ baryons relative to the $\Xiccpp$ baryon, namely $\tau(\Xi_{cc}^{+})/\tau(\Xi_{cc}^{++})=0.22\pm 0.05\pm 0.04$ and $\tau(\Omega_{cc}^{+})/\tau(\Xi_{cc}^{++})=0.52\pm 0.13^{+0.03}_{-0.02}$. 
}

\date{\today}

\begin{document}
\preprint{RBI-ThPhys-2023-9}
\maketitle
\flushbottom

\section{Introduction}

Lifetimes of weakly decaying hadrons containing $c$ and $b$ quarks are typically treated within the heavy quark expansion (HQE) \cite{SV1985} (for review see e.g. \cite{Lenz2014}),  which has the structure of an operator product expansion arranged in inverse powers of the quark mass. For hadrons containing a $b$ quark, the series converges rapidly enough that theoretical predictions of lifetimes can be readily tested against experiment, and in the most recent works, which have included newly-available contributions in the $1/m_b$ and $\als$ expansions, they are found to be in excellent agreement, within uncertainties \cite{Lenz:2022rbq,Gratrex:2023pfn,Cheng:2023voj}. 
In charmed hadrons, however, the HQE only converges slowly, resulting in enhanced sensitivity to uncertainties from the matrix elements and higher-order power corrections.

In our previous work \cite{GMN2022}, we updated the lifetime predictions for weakly decaying singly charmed hadrons. Our results were compatible with the new experimental lifetime for the $\Omc$ baryon \cite{LHCbOmegac2018,LHCb2021Omega0,Belle-II:2022plj} of four times the previous measurements \cite{PDG2018}, albeit with sizeable theoretical uncertainties (see also \cite{Cheng:2023voj}). Some tensions with experiment appeared in lifetime ratios, and in the predictions for $D$ mesons (with our results agreeing with the study of \cite{LenzNote:2021}), but the overall picture is that the heavy quark expansion can consistently describe charmed hadron decays and accommodate experimental data.

In this paper, we extend the analysis of \cite{GMN2022} to the doubly charmed baryons $\Xiccpp,\,\Xiccp,$ and $\Omcc$. Our predictions update those in previous studies \cite{BKLO1998,KLO1998,Likhoded:1999yv,Melic99cc,Melic99cc2,KL2001,CLLW07cc,ChengShi18cc,BLL2018,LL2018} by including the most complete set of available contributions, in particular the Darwin contribution, NLO corrections to two- and four-quark contributions, and subleading (dimension-seven) four-quark contributions. We also provide an estimate of uncertainties arising from renormalisation scale variation and hadronic parameters, and present results for two different charm quark mass schemes.

To date, only the $\Xiccpp$ has been seen  \cite{LHCbXiccpp2017}, and its lifetime determined to be $256(27) \fs$ \cite{LHCbXiccpp2018}. The existence of the other two baryons considered in this work still remains to be confirmed experimentally. While SELEX reported a measurement of $\Xiccp$ \cite{SELEXXiccp2002,SELEXXiccp2004}, with a reported mass of $3.519 \GeV$ and lifetime of $< 33 \fs$, other experimental searches have not confirmed this result \cite{FOCUS2003,BaBarXicc2006,Bellecc2013,LHCbXiccp2013,LHCbXiccp2019,LHCbXiccp2021}, and it has been suggested that the observation at SELEX was a misidentification \cite{Kiselev:2002an}. Likewise, the most recent searches for $\Omcc$ have not reported a detection yet \cite{LHCbOmegacc2021}. 
We find a lifetime for the $\Xiccpp$ consistent with the LHCb result, while the predicted hierarchy $\tau(\Xi_{cc}^{+})  < \tau(\Omega_{cc}^{+}) < \tau(\Xi_{cc}^{++}) $ is in agreement with that in the early studies of \cite{Likhoded:1999yv,FR1989}.

The paper is organised as follows. In section \ref{sec:theory}, we briefly review the heavy quark expansion as applied to doubly charmed baryon decays. In section \ref{sec:matrixelements}, we discuss the nonperturbative inputs, the matrix elements, for the baryons considered in this paper. Our results are presented in section \ref{sec:results}, and we end with conclusions in section \ref{sec:conclusion}. Appendices \ref{app:conventions} and \ref{app:inputs} contain, respectively, the conventions and numerical inputs used in the paper, while in appendix \ref{app:4q} we provide analytic expressions for the leading-order coefficients of four-quark contributions.

\section{Background}

\label{sec:theory}
In this section, we briefly overview the heavy quark expansion (HQE) applied to charm decays. We refer the reader to section 2 of our previous work \cite{GMN2022} for more details, as well as to \cite{Lenz2014,LenzNote:2021}. 

Via the optical theorem, the lifetime of a heavy hadron can be related to the imaginary part of the forward transition operator:
\begin{equation}
\frac{1}{\Lifetime{H}}= \Ga(H) = \frac{1}{2M_H} \langle H| \mathcal{T} |H\rangle\,,\qquad \mathcal{T} = \Im\,\,i\int d^4x\, T\left[\mathcal{H}_{eff}(x)\mathcal{H}_{eff}(0)\right]\,,\label{eq:OpticalTheorem}
\end{equation}
where $\mathcal{H}_{eff}$ is the effective Hamiltonian describing the charged current interactions of the charm quark (e.g.\ \cite{BBL1995})
\begin{align}
    \mathcal{H} =\frac{G_F}{\sqrt{2}}\bigg[&\sum_{q,q'=d,s} V^{\vphantom{\ast}}_{cq}V^\ast_{uq'}\big(C_1(\mu) Q_1^{(qq')}+C_2(\mu) Q_2^{(qq')}\big) \nonumber \\ &{} -V^{\vphantom{\ast}}_{ub}V^\ast_{cb}\sum_{k=3}^{6}C_k(\mu)Q_k
    +\sum\limits_{\substack{q=d,s \\ \ell=e,\mu}}V_{cq}Q^{(q\ell)}\bigg]  +\textrm{h.c.}\,,
    \label{eq:HeffSM}
    \end{align}
where $G_F$ is the Fermi constant, $V_{ab}$ are Cabibbo-Kobayashi-Maskawa (CKM) matrix elements, and the $\Delta C=1$ current-current operators read
\begin{equation}
\begin{split}
    Q_1^{(qq')}&=(\bar{c}^i\gamma_\mu(1-\gamma_5)q^j)(\bar{q}'^j\gamma^\mu(1-\gamma_5)u^i)\,,\\
    Q_2^{(qq')}&=(\bar{c}^i\gamma_\mu(1-\gamma_5)q^i)(\bar{q}'^j\gamma^\mu(1-\gamma_5)u^j)\,,\\
    Q_{\text{SL}}^{(q\ell)}&=(\bar{c}\gamma_\mu(1-\gamma_5)q)(\bar{\ell}\gamma^\mu(1-\gamma_5)\nu_\ell)\,,
    \end{split}
\end{equation}
where $i,j$ are colour indices. The remaining operators $Q_{3\text{-}6}$ denote the penguin operators, which are suppressed by the CKM factor $V^{\vphantom{\ast}}_{ub}V^\ast_{cb}$. Since their Wilson coefficients $C_3\text{-}C_6$ are additionally numerically small (e.g.\ \cite{BBL1995,LenzNote:2021}), we will neglect these contributions in the present paper, as we have done in \cite{GMN2022}.\footnote{Note that $Q_2^{q q'}$ denotes the colour-singlet operator in our convention, following \cite{BBL1995,GMN2022} but opposite to the choice by some other authors, e.g.\ \cite{LenzNote:2021,ChengShi18cc,Melic99cc}, where $Q_1$ is the colour-singlet. }

The right-hand side of \eqref{eq:OpticalTheorem} can then be expanded, using the HQE, in powers of $\Lambda_{QCD}/m_Q$ and $\als$, where $m_Q=m_c$ is the heavy-quark mass and $\Lambda_{QCD}$ is the QCD scale.  

\begin{figure}[th]
    \centering
\includegraphics[scale=0.38,clip=true,trim=100 580 130 30]{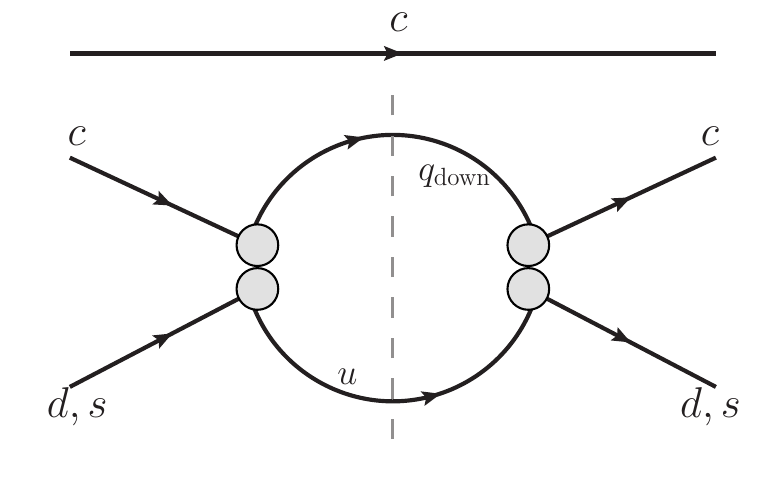}
\includegraphics[scale=0.38,clip=true,trim=100 580 130 30]{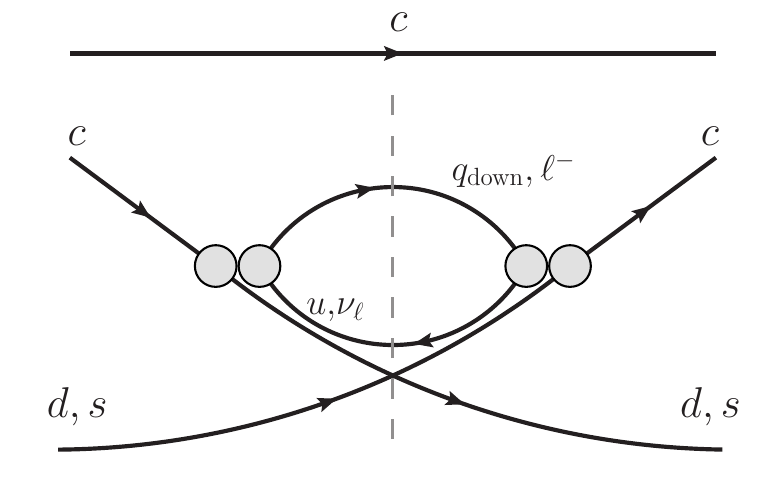}
 \includegraphics[scale=0.38,clip=true,trim=100 580 150 30]{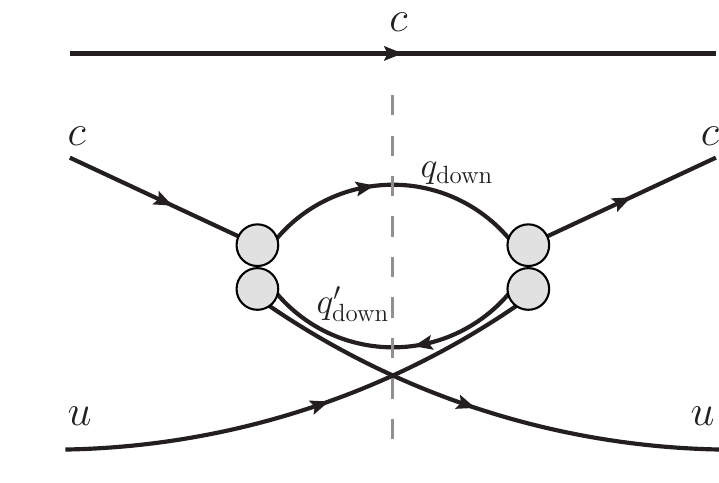}
    \caption{Topologies for four-quark contributions in doubly charmed baryons. From left to right: weak exchange, labeled WE; constructive Pauli interference, labeled $\intp$; destructive Pauli interference, labeled $\intm$. $q_{\rm down}$ and $q_{\rm down}^{\prime}$ in the loop denote $d$ or $s$ quarks.}
    \label{fig:topologies4q}
\end{figure}

 This yields a tower of local operators $\mathcal{O}_i$, ordered by increasing powers of the inverse charm mass,
\begin{equation}
\mathcal{T} = \bigg(\mathcal{C}_3\mathcal{O}_3 +\frac{\mathcal{C}_5}{m_c^2}\mathcal{O}_5 +\frac{\mathcal{C}_6}{m_c^3}\mathcal{O}_6+\dots\bigg)+ 16\pi^2\bigg(\frac{\tilde{\mathcal{C}}_6}{m_c^3}\tilde{\mathcal{O}}_6 + \frac{\tilde{\mathcal{C}}_7}{m_c^4}\tilde{\mathcal{O}}_7 + \dots\bigg)\,,
\label{eq:HQEsystematic}
\end{equation}
where the Wilson coefficients $\mathcal{C}_i$ contain the short-distance physics, analogously to the $C_i$ in \eqref{eq:HeffSM}.\footnote{The absence of the dimension-four operator, suppressed by $\Lambda_{\text{QCD}}/m_{Q}$, was demonstrated in \cite{Chay:1990da,LukeThm}.} 
The operators within the first bracket are each composed of heavy-quark field bilinears, with operators of increasing dimension generated by insertion of covariant derivatives, and will be referred to below as the ``non-spectator'' contributions. 
The terms within the second bracket involve the contributions of four-quark operators. These ``spectator contributions'' are sensitive to the flavour of the light quark in the hadron, and are one-loop enhanced relative to the non-spectator contributions by the factor $16\pi^2$, explicitly exhibited in the above expression. 
Therefore, they can induce significant lifetime splitting effects. There are in fact three distinct topologies for these contributions, referred to as weak exchange (WE), constructive Pauli interference ($\rm{int}^+$), and destructive Pauli interference ($\rm{int}^-$); these are represented in figure \ref{fig:topologies4q}.

The Wilson coefficients $\mathcal{C}_i$ can be calculated perturbatively in powers of the strong coupling constant $\als$,
\begin{equation}
\mathcal{C}_i= \mathcal{C}_i^{(0)}(\mu,\mu_0) + \mathcal{C}_i^{(1)}(\mu,\mu_0) \als(\mu) + \mathcal{C}_i^{(2)}(\mu,\mu_0) \als(\mu)^2 + \dots ,
\label{eq:WilsonCoeffsSchematic}
\end{equation}
where $\mu$ is the renormalisation scale arising from the evolution of the weak Hamiltonian, while $\mu_0$ is an operator factorisation scale. 
Only a few of the $\mathcal{C}_i$ in \eqref{eq:HQEsystematic} are known beyond leading order, see table 4 in our previous work \cite{GMN2022}. Following that work, although some NNLO corrections are known in semileptonic decays, for consistency we include only the available NLO corrections in both semileptonic and nonleptonic contributions. Note that the NLO Darwin contribution, recently computed for $b \to c \ell \nu_\ell$ decays \cite{MMP2021II,Moreno:2022goo}, has not yet been adapted for charm decays, and we therefore also do not include it.

The decay width of a doubly heavy baryon in the HQE can be brought into the form\footnote{For notational consistency with our previous paper \cite{GMN2022}, we use $c_G'$ to distinguish the contribution to the chromomagnetic Wilson coefficient resulting solely from the $1/m_c$-expansion in the soft gluon background field. In terms of the coefficients defined in that paper, $c_G' = c_G - c_3/2$.}
\begin{align}
\Ga(\Barycc) = &~\Ga_0 \bigg[c_3\frac{\langle \Barycc\vert \bar{c}c\vert{\Barycc}\rangle}{2M_{\Barycc}} + \frac{c_G'\muG}{m_c^2} +  \frac{c_\rho\darwin}{m_c^3} + \dots \\
&{}+\frac{16 \pi^2}{2M_{\Barycc}} \bigg( \sum\limits_{i,q}\frac{c_{6,i}^q\langle \Barycc| \Opsix{i}{q}| \Barycc\rangle}{m_c^3} \nonumber
+\sum\limits_{i}\dfrac{c_{7,i}^q\langle \Barycc| \OpsevenP{i}{q}| \Barycc\rangle}{m_c^4}+ \dots \bigg) \bigg]\,,
\label{eq:DecayRateExpansion}
\end{align}
where all CKM contributions are included implicitly in the coefficients $c_i$ in the equation above, and, $\Ga_0$ is the normalisation factor defined by 
\begin{equation}
\Ga_0 = \frac{G_F^2 m_c^5}{192\pi^3} \,. \label{eq:Ga0}
\end{equation}
Because of the leading $m_c^5$ dependence of the decay rate, the charm quark mass has to be precisely defined. This is, however, problematic, owing to the well-known renormalon divergences of the pole mass; for a review, see \cite{Beneke1998,Beneke2021}. To avoid this problem, several  renormalon-free schemes have been introduced, e.g. \cite{Pineda2001,HJS2008,HJLMPSS2017,BSUV1996,FSS2020I,FSS2020II}. However, as many of these schemes were developed initially with the $b$ quark in mind,  it remains unclear how appropriate these are in the case of the charm mass. We will not address this question in this paper, and restrict ourselves to presenting results in terms of the pole and kinetic mass schemes \cite{BSUV1996,FSS2020I,FSS2020II}, following the procedure laid out in section 2.4 of \cite{GMN2022}.

The nonspectator matrix elements are defined as follows \cite{DMT2006}:
\begin{align}
\mukin(\Barycc) &= \frac{-1}{2M_{\Barycc}} \langle \Barycc |\bar{c}_v (iD)^2 c_v | \Barycc \rangle \,, \nonumber \\
\muG(\Barycc)&=\frac{1}{2M_{\Barycc}} \langle \Barycc |\bar{c}_v \frac{1}{2} \sig_{\mu\nu} (g_s G^{\mu\nu}) c_v  |\Barycc \rangle\,, \nonumber \\
\darwin(\Barycc) &= \frac{1}{2M_{\Barycc}}\langle \Barycc |\bar{c}_v (i D_\mu) (i v \cdot D) (i D^\mu) c_v | \Barycc \rangle \,,
\label{eq:nonspecdefns}
\end{align} 
where $c_v$ denotes the phase-redefined QCD field $c_v(x)\equiv e^{im_c v\cdot x}c(x)$. 

Using the definitions in \eqref{eq:nonspecdefns}, one can express the matrix element of the leading $\bar{c}c$ operator as
\begin{align}
\frac{\langle \Barycc | \bar{c}{c}| \Barycc \rangle}{2 M_{{\cal B}_{cc}}} 
=  2 - \frac{\mukin(\Barycc)}{2 m_c^2} + \frac{\muG(\Barycc)}{2 m_c^2}\,,
\label{eq:gcc}
\end{align}
with no further higher order $1/m_c$-corrections~\cite{DMT2006}, while the first term counts the number of heavy quarks in the baryon. 

The dimension-six and dimension-seven four-quark operators, $\Opsix{i}{q}$ and $\OpsevenP{i}{q}$, will be defined in eqs.~\eqref{eq:Dim6BaryonBasis} and \eqref{eq:Dim7BaryonBasis} below.

Considering only the valence contributions, and neglecting doubly Cabibbo-suppressed terms, the leading-order (LO) results for the dimension-six spectator contributions to the decay widths are
\begin{align}
     \tilde{\Ga}_{6, \Xiccpp}&=| V_{cs}|^2 | V_{ud}|^2 \langle  \widehat{\Ga}^u_{6,\intm}(x_s,0)\rangle_{\Xiccpp}+| V_{cs}|^2 | V_{us}|^2\langle \widehat{\Ga}^u_{6,\intm}(x_s,x_s)\rangle_{\Xiccpp} \nonumber \\ & {}
      + | V_{cd}|^2 | V_{ud}|^2 \langle \widehat{\Ga}^u_{6,\intm}(0,0)\rangle_{\Xiccpp}  \,,\label{eq:Xiccppspectator}\pagebreak[2]\\
     \tilde{\Ga}_{6, \Xiccp}&=| V_{cs}|^2 | V_{ud}|^2  \langle \widehat{\Ga}^d_{6,\exc}(x_s,0)\rangle_{\Xiccp} 
      + | V_{cd}|^2 | V_{ud}|^2\left(\langle \widehat{\Ga}^d_{6, \intp}(0,0)\rangle_{\Xiccp}+\langle \widehat{\Ga}^d_{6,\exc}(0,0)\rangle_{\Xiccp} \right) \nonumber\\
     & {} +| V_{cd}|^2\sum_{\ell=e,\mu}\langle \widehat{\Ga}^{d,\text{SL}}_{6,\intp}(x_\ell,0)\rangle_{\Xiccp} \,, \label{eq:Xiccpspectator} \pagebreak[2]\\
     \tilde{\Ga}_{6, \Omcc}&= | V_{cs}|^2 | V_{ud}|^2 \langle \widehat{\Ga}^s_{6,\intp}(0,0)\rangle_{\Omcc} +  | V_{cs}|^2 | V_{us}|^2 \left( \langle \widehat{\Ga}^s_{6,\intp}(x_s,0)\rangle_{\Omcc} + \langle \widehat{\Ga}^s_{6,\exc}(x_s,0)\rangle_{\Omcc}  \right)\nonumber \\
     & {} + | V_{cs}|^2\sum_{\ell=e,\mu}\langle \widehat{\Ga}^{s,\text{SL}}_{6,\intp}(x_\ell,0)\rangle_{\Omcc} \,, 
     \label{eq:Omccspectator}
\end{align}
with the notation referring to the topologies in figure \ref{fig:topologies4q}, and the corresponding expressions given in appendix \ref{app:4q}. NLO corrections to the four-quark contributions are included by following the procedure outlined in section 2.3.2 of \cite{GMN2022}, based on the results of \cite{CFLM2001,FLMT2002,BBGLN2002,LenzRauh2013}. The dimension-seven spectator contributions follow the same pattern and hierarchy as in eqs.~(\ref{eq:Xiccppspectator}-\ref{eq:Omccspectator}).

In the next section, we will discuss the evaluation of the matrix elements appearing in the contributions above.

\section{Matrix elements for doubly charmed baryons}
\label{sec:matrixelements}

Unlike in the case of hadrons containing a single heavy quark, where it is sufficient to deal with both the decay rate and the matrix elements in terms of the HQE, for the case of doubly charmed baryons each matrix element picks up, in effect, two separate contributions. These arise, firstly, from interactions between the two charm quarks, viewed as a diquark in the $S=1$ colour antitriplet state, and secondly from interactions between the $(cc)$ diquark and the remaining light quark in the baryon \cite{EFGM2002,KR2014,BKL2017}.

In order to deal with these, we separately expand the matrix elements in NRQCD and in the HQE, where the former is used for dealing with the charm-charm interactions and the latter for the $(cc)$-$q$ interactions, where we hereafter denote the diquark $(cc)$ by $\diquark$. In order to deal with the charm-charm interactions, we apply the nonrelativistic QCD (NRQCD) expansion of the matrix element of all two quark operators. 

Performing the standard Foldy-Wouthuysen transformation of the charm quark field results in the following NRQCD expansion of the $\bar{c}c$ operator \cite{BB1996} in terms of the nonrelativistic two-component field $\Psic$:
\begin{align}
    \bar{c}c & = \Psipc \Psic - \frac{1}{2m_c^2} \Psipc (i \vec{D})^2 \Psic + \frac{3}{8m_c^4} \Psipc (i \vec{D})^4 \Psic \nonumber \\
    & {} - \frac{1}{2m_c^2} \Psipc g_s \vec{\sig} \cdot \vec{B} \Psic - \frac{1}{4m_c^3} \Psipc g_s (\vec{{\cal D}} \cdot \vec{E} )\Psic + \dots
    \label{eq:QQNRQCDexp}  
\end{align}
at $\mathcal{O}(1/m_c^4)$, ie at $O(v_c^7)$ in the counting in terms of nonrelativistic $c$-quark velocity $v_c$. 
The velocity scaling rules in NRQCD \cite{LMNMH1992} 
\begin{align}
    \Psic \sim (m_c v_c)^{3/2}, \quad\quad \vec{D} \sim m_c v_c , \quad\quad g_s \vec{B} \sim m_c^2 v_c^4, \quad\quad g_s \vec{E} \sim m_c^2 v_c^3\,,
    \end{align}
indicate that up to $O(v_c^7)$ we must keep the $\vec{D}^4$ term, and that the Darwin term is, despite being $1/m_c$ suppressed, of the same order in $v_c$ counting as the chromomagnetic term.  

In addition, we will make use of the NRQCD expansion of the chromomagnetic operator
\begin{align}
\bar{c} \frac{1}{2} \sigma_{\mu\nu}G^{\mu\nu} c = -    \Psipc \, g_s \, \vec{\sig} \cdot \vec{B} \Psic - \frac{1}{2 m_c}   \Psipc \, g_s \, (\vec{{\cal D}} \cdot \vec{E}) \, \Psic + \dots
\label{eq:chromoNR}
\end{align}
In the above expressions, one differentiates between the covariant derivative $\vec{D}$ in the fundamental representation, and the covariant derivative $\vec{{\cal D}}$ in the adjoint representation. Furthermore, we use the notation $(\vec{{\cal D}} \cdot \vec{E})$ to denote that the derivative acts inside the bracket only; the remaining conventions can be found in appendix \ref{app:conventions}.


\subsection[Matrix element of the kinetic operator]{Matrix element of the kinetic operator $\mu_{\pi}^2$}

To extract the kinetic parameters, arising from the structures $\Psipc (i \vec{D})^2 \Psic$ and $\Psipc (i \vec{D})^4 \Psic$,
we use the nonrelativistic picture and the observation from the potential models (from where the average kinetic energy of a bound system is derived by using the virial theorem) that the quark kinetic energy practically does not depend on the quark content of the system, but is determined by its colour structure. We first estimate the matrix element
\begin{equation}
    \frac{\langle \Barycc \vert \Psi_c^\dagger (i\vec{D})^2\Psi_c\vert \Barycc \rangle}{2M_{\Barycc}}=2m_c^2\langle v_c^2\rangle\,,
\end{equation}
where the factor of 2 arises from the presence of two charm quarks, and $\langle v_c^2\rangle$ is the mean squared three-velocity of a single charm quark in the baryon. The latter can be estimated as $\langle v_c^2\rangle=\langle v_c({\cal D})^2\rangle+\langle v_{\cal D}^2\rangle$, where $v_c({\cal D})$ is the velocity of a charm quark within the diquark, and $v_{\cal D}$ is the velocity of the diquark. 
Then, by following  \cite{KLO1998}, we express the total kinetic energy $T$ in the baryon's rest-frame as 
\begin{align}
    T = \frac{m_{\cal D}  v_{\cal D}^2  }{2} + \frac{m_{q}  v_q^2  }{2} \,,
\end{align}
where $m_{\cal D} \simeq 2m_c$ is the mass of the diquark, and $v_q$ is the velocity of the light quark. Using this expression, and applying conservation of the momentum in the baryon's rest frame, the mean squared velocity of the diquark is 
\begin{align}
   \langle  v_{\cal D}^2 \rangle  = \frac{m_q T}{2 m_c^2 + m_c m_q}\,.
\end{align}
On the other hand, by expressing the average kinetic energy of the diquark, accounting for a factor 1/2 stemming from the color factor difference between potentials of colour antitriplet-triplet (bound in a singlet state), and colour triplet-triplet (bound in an antitriplet state) systems, as  
\begin{align}
\frac{T}{2}= \frac{1}{2} m_{c_1} v_{c_1}^2({\cal D}) + \frac{1}{2} m_{c_2} v_{c_2}^2({\cal D})
\end{align}
and, assuming $\langle v_{c_1}({\cal D}) \rangle = \langle v_{c_2}({\cal D}) \rangle = \langle v_c({\cal D}) \rangle$,
one can extract the mean squared velocity of the $c$-quark in the diquark as \cite{KLO1998}
\begin{align}
  \langle  v_c^2({\cal D})\rangle  = \frac{T}{2 m_c} \,.
\end{align}
Combining both formulas, we finally obtain the mean squared velocity of a $c$-quark in a doubly charmed baryon:
\begin{align}
    \langle v_c^2 \rangle = \frac{T}{2 m_c} \left ( 1 + \frac{m_q}{ m_c+ m_q/2}  \right ) \,,
\end{align}
where it is evident that the dominant contribution comes from the motion of the charm quark within the diquark. 

Following \cite{KLO1998,KL2001}, which has subsequently been applied in \cite{Melic99cc,ChengShi18cc}, we have finally the expression 
\begin{align}
     \frac{\langle\Barycc\vert \Psi_c^\dagger (i\vec{D})^2\Psi_c\vert \Barycc\rangle}{2M_{\Barycc}} = 2 m_c^2 \left( \frac{T}{2m^\Bary_c} + \frac{m^\Bary_q T}{2(m^\Bary_c)^2 + m^\Bary_c m^\Bary_q}\right)  \,,
    \label{eq:kineticNRQCD}
\end{align}
As was done implicitly in \cite{KLO1998}, and made clear in \cite{Melic99cc}, the quark masses arising from the matrix element should be treated as \textit{constituent} masses for consistency, with modern values, taken from \cite{KR2014}, given in eqs.~\eqref{eq:mqeffmes} and \eqref{eq:mqeffbary}. For the kinetic energy, we use the value $T=0.4\,\text{GeV}$ from \cite{GKLT1994,KLO1998}, leading to the value $\langle v_c^2 \rangle=0.14$, to which we assign a $30\%$ uncertainty. 

 Finally, the $\Psipc (i \vec{D})^4 \Psic$ contribution can be approximated in terms of the squares of the $\Psipc (i \vec{D})^2 \Psic$ contribution, which is to say that
\begin{equation}
    \frac{1}{2M_{\Barycc}} \langle \Barycc | \Psipc (i \vec{D})^4 \Psic | \Barycc \rangle \approx 2 m_c^4 \langle v_c^2 \rangle ^2 
     \label{eq:kineticNRQCD2}
\end{equation}
with the same inputs as before. 
Again, the factor of two accounts for two $c$-quarks in the doubly charmed baryon.

\subsection[Matrix element of the chromomagnetic operator]{Matrix element of the chromomagnetic operator $\mu_G^2$}

To estimate the matrix element $\muG$, we can use the HQE for the baryon mass \cite{FN92I,FN92II,Neubert1996},
\begin{equation}
M_\Barycc = 2m_c + \bar{\La} + \frac{\mukin(\Barycc)}{2m_c} - \frac{\muG(\Barycc)|_{\diquark\textrm{-}q}}{2m_c}- \frac{\muG(\Barycc)|_{c\textrm{-}c}}{2m_c} + \mathcal{O}\left(\frac{1}{m_c^2}\right) \,.
\label{eq:mHexpand}
\end{equation}
where $\bar{\La} \sim 0.5 \GeV$ for charmed hadrons, and all parameters in the expansion are formally independent of the heavy quark mass. For doubly charmed baryons, $\muG$ arises from two distinct sources: the diquark-quark chromomagnetic interaction, and the chromomagnetic interactions between the charm quarks within the diquark. 
In the constituent picture, these can be written, emphasising the dependence on spin, as
\begin{align}
\muG(\Barycc)|_{\diquark\textrm{-}q} &= d_{\Barycc} \lambda_2(\Barycc)|_{\diquark\textrm{-}q} \,, \quad
    \muG(\Barycc)|_{c\textrm{-}c} =  d_{\diquark} \lambda_2(\Barycc)|_{c\textrm{-}c} \,,
\end{align}
respectively,  where $d_{\Barycc}$ and $d_{\diquark}$ denote the spin factors
\begin{align}
\label{eq:dHdef}
d_\Barycc &= -2 \left(S_{\cal B} (S_{\cal B}+1) - S_{\diquark} (S_{\diquark} +1 ) - S_l(S_l+1)\right) \,, \nonumber \\
d_{\diquark} & = -2 \left(S_{\diquark} (S_{\diquark} +1 ) - 2 S_c (S_c + 1) \right) \,.
\end{align}
The standard procedure of extraction of $\muG$ is to take the hyperfine splitting between the masses of spin-$1/2$ baryon $\Barycc$ and its spin-$3/2$ $\Baryccstar$ partner. 
Since the spin of the $(cc)$ diquark does not change between the $\Barycc$ and $\Baryccstar$ states, the contribution of $\muG(\Barycc)|_{c\textrm{-}c}$ to the mass expansion \eqref{eq:mHexpand} above cancels in the mass difference $M_{\Baryccstar} - M_{\Barycc}$, so that the $\muG(\Barycc)|_{\diquark\textrm{-}q}$ part can be  extracted. The relevant values for our purposes are $d_{\Barycc} = 4$, $d_{\Baryccstar} = -2$, and $d_{\diquark} = -1$, since the diquark ${\diquark}$ is in a spin-1 state in both the spin-1/2 and spin-3/2 baryons. One obtains \begin{equation}
\label{eq:muGcqdef1}
\left.\muG(\Barycc)\right|_{\diquark\textrm{-}q} = d_{\Barycc} 2 m_c \frac{M_{\Baryccstar} - M_{\Barycc}}{d_{\Barycc} - d_{\Baryccstar}} = \frac{4}{3}m_c \left(M_{\Baryccstar} - M_{\Barycc}\right)\,, 
\end{equation}
which agrees with \cite{ChengShi18cc,Melic99cc}, up to their differing presentation of factors of two.\footnote{In \cite{KLO1998}, there is an apparent sign error in the same contribution.}

To calculate the remaining $\muG$ piece arising from the $c$-$c$ interaction, we evaluate the $c$-$c$ contribution by taking the matrix element of the nonrelativistic expression \eqref{eq:chromoNR} in the constituent model, and obtain
\begin{align} 
\left.\frac{\langle \Barycc |\Psipc g_s\vec{\sig}\cdot \vec{B} \Psic | \Barycc \rangle}{2M_{\Barycc}} \right|_{c\textrm{-}c}   = 2 \cdot \left( - \frac{4}{3} \right) \frac{1}{m_c} g_s^2 \, \langle t_1^a t_2^a \rangle \, \langle \vec{s}_1 \cdot \vec{s}_2 \rangle \,|\psi_{cc}(0) |^2 \,,
      \label{eq:chromocc1}
\end{align}
where $t_{1,2}^a$  and $\vec{s}_{1,2}$ are the color matrices and the spin vectors of the two $c$-quarks in the diquark, and $\psi_{cc}(0)$ is the wave function of the diquark at the origin \cite{KLO1998,KL2001,Melic99cc}. For the spin-1 diquark we have
$\langle t_1^a t_2^a \rangle  = -2/3$ and   $\langle \vec{s}_1 \cdot \vec{s}_2 \rangle = 1/4$, so that
\begin{equation}
 \left. \frac{\langle \Barycc |\Psipc g_s\vec{\sig}\cdot \vec{B} \Psic | \Barycc \rangle}{2M_{\Barycc}} \right|_{c\textrm{-}c}   = \frac{4}{9} \frac{g_s^2}{m_c} |\psi_{cc}(0)|^2 \,.
  \label{eq:chromomatelcc}
\end{equation}
Using the equation of motion for the gluon field strength \eqref{eq:EOM} in the nonrelativistic limit, we can relate the remaining nonrelativistic Darwin term in \eqref{eq:chromoNR} to the diquark wave function as 
\begin{align}
\left.\frac{ \langle \Barycc |\Psipc \, g_s (\vec{{\cal D}} \cdot \vec{E})\, \Psic |\Barycc \rangle }{2M_{\Barycc}} \right|_{c\textrm{-}c} =  - g_s^2 \frac{ \langle \Barycc| (\Psipc \, t^a \Psic )(\Psipc \, t^a \Psic ) | \Barycc \rangle }{2M_{\Barycc}}
 = \frac{4}{3} g_s^2 |\psi_{cc}(0)|^2 \,.
\end{align}
Therefore, the final expression for the matrix element of the chromomagnetic operator for doubly heavy baryons reads:
\begin{align}
\muG(\Barycc) & = \muG(\Barycc)|_{\diquark\textrm{-}q} + \muG(\Barycc)|_{c\textrm{-}c} \nonumber \\  
& =2 \bigg [ \frac{2}{3}\,m_c\, \left(M_{\Baryccstar} - M_{\Barycc}\right) -\frac{2}{9} \frac{g_s^2}{m_c} |\psi_{cc}(0)|^2 - \frac{1}{3} \frac{g_s^2}{m_c} |\psi_{cc}(0)|^2 \bigg ]\,.
\end{align}

The diquark wave function $|\psi_{cc}(0)|$ is the nonrelativistic radial wave function at the origin for a $(cc)$ diquark system.  Its extraction is not straightforward, since it cannot be related to a physical decay constant, as is the case for $c\bar{c}$ systems, where
$|\psi_{c\bar{c}}(0)|^2 = 1/12 \, f_{J/\Psi}^2  M_{J/\psi}$.
A naive assumption $|\psi_{cc}(0)|^2 = |\psi_{c\bar{c}}(0)|^2$ is also not a satisfying choice, given that the binding potentials of $(cc)$ and $(\bar{c}c)$ systems differ by an overall colour-related factor.\footnote{Using this assumption to extract the value of $|\psi_{c\bar{c}}(0)|^2$ from the $J/\psi$ decay constant, $f_{J/\psi} = 418(8)(5) \, {\rm MeV}$ \cite{BDKMS2013}, we obtain 
\begin{align}
\vert\psi_{cc}(0)\vert^2 = \vert\psi_{c\bar{c}}(0)\vert^2 = 0.045 \, {\rm GeV}^3 \,. \nonumber 
\end{align}
Another option is to use the simple dimensional scaling relation that takes into account the ratio of the colour factors in the binding potentials of the $(cc)$ and $(\bar{c}c)$ systems \cite{FLSW1993, G2008}, which leads to
\begin{align}
  |\psi_{cc}(0)|^2 = \frac{1}{8} |\psi_{c\bar{c}}(0)|^2 = 5.6 \cdot 10^{-3}  \, {\rm GeV}^3\,,\nonumber
\end{align} 
which differs by about an order of magnitude from the value obtained in the potential models.
}
For the present analysis, we use the value obtained from the fits to the physical charm hadron masses using the potential models in \cite{BDGGN1994}
\begin{equation}
\vert\psi_{cc}(0)\vert^2=0.02\,\text{GeV}^3\,,
\label{eq:psiCC}
\end{equation}
to which we also assign a conservative $30\%$ model uncertainty.

Combining all the above contributions together leads to our expression for the dimension-three matrix element:  
\begin{align}
\frac{\langle \Barycc | \bar{c}c | \Barycc \rangle}{2M_{\Barycc}} = 2 & \bigg [1 - \frac{1}{2}\langle v_c^2 \rangle +  \frac{3}{8}\langle v_c^2 \rangle^2 + \frac{1}{3} \frac{M_{\Baryccstar} - M_{\Barycc} }{m_c} \nonumber\\
&- 4\pi \als  \, \frac{1}{9}  \frac{ |\psi_{cc}(0)|^2}{m_c^3} -  4\pi \als \, \frac{1}{6}  \frac{|\psi_{cc}(0)|^2}{m_c^3}   \bigg ]\,.\label{eq:dimension-3Expansion}
\end{align}
Note that the apparent discrepancy in the overall factor of the above matrix element that is missing in some of the previous literature \cite{KLO1998,Melic99cc,ChengShi18cc} can be attributed to a difference in presentation style, wherein  only one charm quark in the matrix elements was accounted for, but the factor of two was instead included in the initial formula for the operator product expansion in eq.~\eqref{eq:OpticalTheorem}. In addition, the last two terms in (\ref{eq:dimension-3Expansion}) are, in \cite{ChengShi18cc}, smaller by a factor of two, while we agree with \cite{KLO1998,Melic99cc}.

The sizes of different contributions can be illustrated numerically by evaluating the above expression for the central values of $T = 0.4 \GeV$ \cite{GKLT1994,KLO1998} and the value for $|\psi_{cc}(0)|^2$ in \eqref{eq:psiCC}, with $\als(1.5\,\text{GeV})=0.35$:
\begin{align}
   \frac{\langle \Barycc | \bar{c}c | \Barycc \rangle}{2M_{\Barycc}} = 2 - 0.14 + 0.02 + 0.04 - 0.006 - 0.009\, , 
\end{align}
where the ordering of the numerical values follows the ordering of the terms in the formula \eqref{eq:dimension-3Expansion}. 

\subsection{Matrix elements of the operators involving spectator interactions}

Turning to the four-quark matrix elements, we will apply the nonrelativistic constituent quark model (NRCQM), in which the matrix elements, for the basis of dimension-six operators
\begin{alignat}{3}
\Opsix{1}{q} &= (\bar{c}_i \ga_\mu(1-\ga_5) q_i) (\bar{q}_j \ga^\mu(1-\ga_5) c_j) \,, \quad
& \Opsix{2}{q} &= (\bar{c}_i (1-\ga_5) q_i) (\bar{q}_j (1+\ga_5) c_j) \,,\nonumber \\
\Opsixt{1}{q} &= (\bar{c}_i \ga_\mu(1-\ga_5) q_j) (\bar{q}_j \ga^\mu(1-\ga_5) c_i) \;,\quad
&\Opsixt{2}{q} &= (\bar{c}_i (1-\ga_5) q_j) (\bar{q}_j (1+\ga_5) c_i) \,, \label{eq:Dim6BaryonBasis}
\end{alignat} 
where $i,j,$ are colour indices, can be expressed in terms of the baryon wave functions as \cite{ChengShi18cc} 
\begin{align}
\langle \Opsix{1}{q} \rangle_{\Barycc} = \cfrac{\langle \Barycc| \Opsix{1}{q}| \Barycc\rangle}{2M_{\Barycc}} &= - 6\,|\Psi_{cq}^{\Barycc}(0)|^2 \,, \qquad 
\langle \Opsix{2}{q} \rangle_{\Barycc} = \cfrac{\langle \Barycc| \Opsix{2}{q}| \Barycc\rangle}{2M_{\Barycc}} = - \,|\Psi_{cq}^{\Barycc}(0)|^2 \,, \nonumber \\
\langle \Opsixt{1}{q} \rangle_{\Barycc} = \cfrac{\langle \Barycc| \Opsixt{1}{q}| \Barycc\rangle}{2M_{\Barycc}} &=  6\tilde{B}\,|\Psi_{cq}^{\Barycc}(0)|^2 \,, \qquad 
\langle \Opsixt{2}{q} \rangle_{\Barycc} =\cfrac{\langle \Barycc| \Opsixt{2}{q}| \Barycc\rangle}{2M_{\Barycc}} =  \tilde{B}\,|\Psi_{cq}^{\Barycc}(0)|^2 \,,
\label{eq:matels4q}
\end{align}
where $\tilde{B}=1$ in the constituent quark limit \cite{NS1996}, which is taken as exact in the present analysis. Note that the matrix elements are only nonzero when the quark in the operator matches the valent quark in the baryon.\footnote{In principle, non-valent contributions, via so-called `eye contractions', will also contribute, but these have not been computed for baryons \cite{GMN2022,Gratrex:2023pfn}. They can, however, be expected to be subleading, as was observed in mesons \cite{KLR2021}, and we therefore neglect them in this study.} 

Following the mass formula of de Rujula, Georgi, and Glashow \cite{GGR1975}, we relate the wave functions in \eqref{eq:matels4q} to the baryon hyperfine splittings. 
Normalising to the equivalent hyperfine splitting of charmed mesons, in order to reduce uncertainties from the constituent quark masses and the scale of $\als$, leads to the relation
\begin{equation}
    \left|\Psi^{\Barycc}_{cq}(0) \right|^2 = \frac{4}{3} y_q \frac{M_{\Baryccstar} - M_{\Barycc}}{M_{D_q^*} - M_{D_q}} \vert\Psi^{D_q}_{cq}(0) \vert^2 \,,
    \label{eq:wavefnratio}
\end{equation}
where $\vert\Psi^{D_q}_{cq}(0) \vert^2 = (1/12)\, f_{D_{q}}^2 m_{D_q}$ is the meson wave function, and $y_q$ is the ratio of constituent quark masses between baryons and mesons, $(m_c^\Bary m_q^\Bary)/(m_c^\Mes m_q^\Mes)$; using the values for quark masses quoted in eqs.~\eqref{eq:mqeffmes} and \eqref{eq:mqeffbary} \cite{KR2014} leads to
\begin{eqnarray}
y_u =y_d \equiv y = \frac{m_c^\Bary m_q^\Bary}{m_c^\Mes m_q^\Mes} \simeq 1.20\,, \qquad \qquad
y_s = \frac{m_c^\Bary m_s^\Bary}{m_c^\Mes m_s^\Mes} \simeq 1.15 \,.
\label{eq:yqsvals}
\end{eqnarray}
However, with the exception of $m_{\Xiccpp}$, the masses of doubly charmed baryons have yet to be determined experimentally. We are therefore obliged to turn to a model estimate of their values. Numerous studies exist, and we refer the reader to the tables in \cite{Soto:2020pfa,Soto:2021cgk,Yu:2022lel} for a comprehensive overview. In the heavy quark limit $m_c \to \infty$, there is the relation \cite{SW1990,BRV2005}
\begin{equation}
    \frac{4}{3} \frac{M_{\Baryccstar} - M_{\Barycc}}{M_{D_q^*} - M_{D_q}} \to 1 \,,
\end{equation}
although it can be expected that this will only approximately hold in the charm sector. In this study, we will take the values for masses and hyperfine splittings from the lattice QCD computation in \cite{BDMO2014}: 
\begin{equation}
    M_{\Xiccstar}-M_{\Xicc}=82.8\pm 9.2\,\text{MeV}\,,\qquad \qquad M_{\Omccstar}-M_{\Omcc}=83.8\pm 5.5\,\text{MeV}\,,
\end{equation}
where we combined the reported statistical and systematic uncertainties in quadrature. This leads to the values for the matrix elements given in table~\ref{tab:matels4qvalues}.

\begin{table}[ht]
\renewcommand{\arraystretch}{1.8}
    \centering
    \begin{tabular}{|c|c|}
    \hline
    Matrix element & value/$\GeV^3$ \\ \hline
   $\langle O_1^u \rangle_{\Xiccpp} = \langle O_1^d \rangle_{\Xiccp}$      & $-0.040 \pm 0.003 \pm 0.01$  \\
   $\langle O_1^s \rangle_{\Omcc}$      & $-0.055 \pm 0.004 \pm 0.02$ \\ \hline
    \end{tabular}
    \caption{Values of the matrix elements $\langle O_1^q \rangle_{\Barycc} = \langle {\Barycc} | O_1^q | {\Barycc} \rangle/ (2m_{\Barycc}) $ using the baryon hyperfine splitting values from \cite{BDMO2014}, with remaining parameters given in appendix \ref{app:inputs}. The first uncertainty arises from the input parameters, while the second one results from assigning a $ 30\%$ error to account for inherent model uncertainties.
    The other four-quark matrix elements can be inferred from the relations in eq.\ \eqref{eq:matels4q}. }
    \label{tab:matels4qvalues}
\end{table}

For the dimension-seven matrix elements, where the operator basis is \cite{LenzRauh2013,ChengShi18cc,LenzNote:2021,GMN2022}
\begin{align}
\OpsevenP{1}{q} &=  m_q (\bar{c}_i (1-\ga_5) q_i) (\bar{q}_j (1-\ga_5) c_j) \,, \nonumber \\ 
\OpsevenP{2}{q} &= \frac{1}{\mQheavy}(\bar{c}_i \Dleft_\rho \ga_\mu(1-\ga_5) D^\rho q_i) (\bar{q}_j  \ga^\mu(1-\ga_5) c_j) \,, \nonumber \\
\OpsevenP{3}{q} &= \frac{1}{\mQheavy}(\bar{c}_i \Dleft_\rho (1-\ga_5) D^\rho q_i) (\bar{q}_j  (1+\ga_5) c_j) \,, \nonumber \\
\OpsevenPt{1}{q} &=  m_q (\bar{c}_i (1-\ga_5) q_j) (\bar{q}_j (1-\ga_5) c_i) \,, \nonumber \\
\OpsevenPt{2}{q} &= \frac{1}{\mQheavy}(\bar{c}_i \Dleft_\rho \ga_\mu(1-\ga_5) D^\rho q_j) (\bar{q}_j  \ga^\mu(1-\ga_5) c_i) \,, \nonumber \\
\OpsevenPt{3}{q} &= \frac{1}{\mQheavy}(\bar{c}_i \Dleft_\rho (1-\ga_5) D^\rho q_j) (\bar{q}_j  (1+\ga_5) c_i) \,, \label{eq:Dim7BaryonBasis}
\end{align}
we will follow \cite{GMN2022} to relate these to dimension-six matrix elements by the scaling relations (with the definitions $\langle \OpsevenP{i}{q} \rangle_{\Barycc} = \langle \Barycc | \OpsevenP{i}{q} | \Barycc \rangle/ (2M_{\Barycc}) $)
\begin{align}
    \langle \OpsevenP{1}{q} \rangle_{\Barycc} = m_q \langle \Opsix{2}{q} \rangle_{\Barycc} \,, \quad  \langle \OpsevenP{2}{q} \rangle_{\Barycc} = \Lambda_{\rm QCD} \langle \Opsix{1}{q} \rangle_{\Barycc} \,, \quad \langle \OpsevenP{3}{q} \rangle_{\Barycc} = \Lambda_{\rm QCD} \langle \Opsix{2}{q} \rangle_{\Barycc} \,,
\end{align}
where we used $\Lambda_{\rm QCD} = 0.33\,\text{GeV}$ \cite{HS2017} for the central values. The matrix elements from the remaining three dimension-seven operators follow the colour-antisymmetry relation $\langle \OpsevenPt{i}{q} \rangle_{\Barycc} = - \tilde{B} \langle \OpsevenP{i}{q} \rangle_{\Barycc}$, with $\tilde{B} = 1$ in the valence quark approximation \cite{NS1996}, which we again take to be exact. 

\subsection[Matrix element of the Darwin operator]{Matrix element of the Darwin operator $\rho_D^3$}
The coefficient of the Darwin term in the decay rate was only recently computed in \cite{MMP2020,LPR2020,Moreno2020,LenzNote:2021}, since at the level of ${\cal O}(1/m_c^3)$ the Darwin operator starts to mix with the four-quark operators and computational difficulties arise. For the corresponding matrix element, we will apply the equation of motion \eqref{eq:EOM} for the gluon field strength to relate the contribution from the $\diquark\text{-}q$ interaction to the four-quark matrix elements, following \cite{GMN2022,Gratrex:2023pfn}, giving
\begin{equation}
    2M_{\Barycc}\rho_D^3=
    g_s^2 \sum\limits_{q=u,d,s}\langle \Barycc| \big(-\frac{1}{8}O^q_1+\frac{1}{24}\tilde{O}^q_1+\frac{1}{4}O^q_2-\frac{1}{12}\tilde{O}^q_2\big)| \Barycc\rangle\,+\mathcal{O}(1/m_c)\,,\label{eq:Darwin4qbar}
\end{equation}
where only the terms that survive in the valence quark approximation are kept in the sum. 

The remaining piece, from the interaction between the two charm quarks, can be estimated in like manner to what was done for the chromomagnetic term in \eqref{eq:chromomatelcc}, with the result
\begin{equation}
 \rho_D^3(\Barycc)|_{c\rm{-}c} = \frac{1}{2} \frac{\langle \Barycc |\Psi^\dag g_s(\vec{\cal D}\cdot \vec{E}) \Psi | \Barycc \rangle}{2M_{\Barycc}}   = \frac{2}{3} g_s^2 |\Psi_{cc}(0)|^2 \,.
  \label{eq:rhoDmatelcc}
\end{equation}
Combining the contributions \eqref{eq:Darwin4qbar} and \eqref{eq:rhoDmatelcc}, using the relations between the matrix elements from eq.\ \eqref{eq:matels4q}, and inserting the values in table~\ref{tab:matels4qvalues}, gives the values 
\begin{equation}
  \rho_D^3(\Xiccpp,\Xiccp)=0.08 \pm 0.02 \GeV^3\,,\qquad \qquad \rho_D^3(\Omcc) = 0.09 \pm 0.02 \GeV^3 \,, 
  \label{eq:matelDarwinvalues}
\end{equation}
where the dominant contribution arises from the charm-charm interaction \eqref{eq:rhoDmatelcc}, and the uncertainty in \eqref{eq:matelDarwinvalues} arises from the $\vert\psi_{cc}(0)\vert^2$ model uncertainty. 

\section{Results and discussion}
\label{sec:results}
In this section, we present our predictions for the lifetimes of doubly charmed baryons and their lifetime ratios. We estimate the scale uncertainties by varying the scale $\mu$ in the range $(1\GeV, 3\GeV)$, while the central values correspond to fixing $\mu = 1.5\GeV$. We also take into account hadronic uncertainties, outlined in the previous section, and (smaller) parametric uncertainties. For the values of the charm quark mass in the pole and kinetic mass scheme, we follow our previous work \cite{GMN2022}.

The predicted lifetime ratios
are obtained using 
\begin{equation}
\frac{\tau(H_2)}{\tau(H_1)} 
=\frac{1}{1 + \left[\Gamma(H_2) - \Gamma(H_1)\right]^{\rm theory} \tau(H_1)^{\rm exp}}\,,
\label{Eq: RatioDef}
\end{equation}
where for the experimental normalisation we use the value of the $\Xiccpp$ lifetime measured by the LHCb Collaboration \cite{LHCbXiccpp2018},
\begin{equation}
\tau(\Xiccpp)^{\rm LHCb}=0.256^{+0.024}_{-0.022}\pm 0.014\,{\rm ps}\,.
\label{eq:LHCbtXiccpp}
\end{equation}
Expressing the lifetime ratio in this way results in the cancellation of the universal non-spectator contributions, which therefore leads to a reduction in the theoretical uncertainties.

Our predictions for the lifetimes and ratios are given in table~\ref{tab:results}.  These results are consistent with the hierarchy
\begin{equation}
\tau(\Xiccp) < \tau(\Omcc) < \tau(\Xiccpp) \,. 
\label{eq:ourhierarchy}
\end{equation}
This hierarchy was first predicted in  \cite{FR1989}, and follows from a consideration of the relative sizes and signs of the four-quark contributions.  In terms of the topologies in figure \ref{fig:topologies4q}, the $\Xiccp$ decay width can be expected to pick up a large, positive exc contribution, while the $\Xiccpp$ picks up a small (negative) $\intm$ contribution. The dominant $\intp$ contribution to $\Omcc$ is also positive, driving its lifetime somewhat lower than that of the $\Xiccpp$. While previous estimates of the lifetime hierarchy were based solely on the LO spectator contributions, we find the hierarchy to be robust even when including radiative corrections to the dimension-six operators together with the LO dimension-seven contributions.

 We illustrate the sizes of different contributions to the overall decay widths using the central values of the inputs, in the kinetic mass scheme, in the following numerical formulas:
\begin{align}
    \Gamma(\Xiccpp)/{\rm ps^{-1}} &= ( \underbrace{2.22}_{\rm LO} + \underbrace{1.14}_{\rm NLO} )  - 0.14 \frac{\mukin}{0.5 \GeV^2} + 0.003 \frac{\muG}{0.13 \GeV^2}  + 0.35 \frac{\darwin}{0.08 \GeV^3} \nonumber \\ & {} + ( \underbrace{-1.43}_{\rm LO} + \underbrace{0.15}_{\rm NLO} ) \frac{\langle O_1^u \rangle}{-0.04 \GeV^3} - 0.05\,\frac{1}{-8.5\cdot 10^{-4}\,\text{GeV}^4} \underbrace{\langle P_1^u \rangle}_{=0} + 0.82 \frac{\langle P_2^u \rangle}{-0.01 \GeV^4} \,,\label{Eq:Xiccplpl-breakdown}
\end{align}
\begin{align}
    \Gamma(\Xiccp)/{\rm ps^{-1}} &= ( \underbrace{2.22}_{\rm LO} + \underbrace{1.14}_{\rm NLO} )  - 0.14 \frac{\mukin}{0.5 \GeV^2} + 0.003 \frac{\muG}{0.13 \GeV^2}  + 0.35 \frac{\darwin}{0.08 \GeV^3} 
 \nonumber \\ & {} + ( \underbrace{7.27}_{\rm LO} + \underbrace{3.17}_{\rm NLO} ) \frac{\langle O_1^d \rangle}{-0.04 \GeV^3} + 7\cdot 10^{-3} \frac{1}{-8.5\cdot 10^{-4}\,\text{GeV}^4}\underbrace{\langle P_1^d \rangle}_{=0} + 3.25 \frac{\langle P_2^d \rangle}{-0.01 \GeV^4} \,,\label{Eq:Xiccpl-breakdown}
\end{align}
\begin{align}
    \Gamma(\Omcc)/{\rm ps^{-1}} &= ( \underbrace{2.22}_{\rm LO} + \underbrace{1.14}_{\rm NLO} )  - 0.14 \frac{\mukin}{0.5 \GeV^2} + 0.003 \frac{\muG}{0.13 \GeV^2}  + 0.39 \frac{\darwin}{0.09 \GeV^3} \nonumber \\ & {} + ( \underbrace{5.37}_{\rm LO} + \underbrace{0.11}_{\rm NLO} ) \frac{\langle O_1^s \rangle}{-0.05 \GeV^3}  + 0.13 \frac{\langle P_1^s \rangle}{-8.5 \cdot 10^{-4} \GeV^4} \nonumber \\
    & {} - 2.50 \frac{\langle P_2^s \rangle}{-0.02 \GeV^4} + 0.18 \cdot 10^{-4}\frac{\langle P_3^s \rangle}{-0.003 \GeV^4}\,.\label{Eq:Omcc-breakdown}
\end{align}
Several observations regarding the relative magnitudes of the contributions are now in order. We first note that we have neglected the masses of up- and down-quarks, resulting in the vanishing of the dimension-seven matrix elements $\langle P_1^{u,d} \rangle$; see eq.~\eqref{eq:Dim7BaryonBasis}. Furthermore, the Darwin term exhibits the largest impact among the power-suppressed two-quark contributions. Its effect on total decay rates is, however, mitigated by cancellation with the kinetic term. Regarding the spectator contributions, we note the sizeable effects from NLO corrections to the dimension-six terms. These, however, contribute in the same direction as the LO terms, except for the relatively small opposite-sign contribution in the case of negative Pauli interference in $\Gamma(\Xiccpp)$. The small NLO correction to the dimension-six contribution for the $\Omcc$ baryon in the kinetic scheme, evident from eq.~\eqref{Eq:Omcc-breakdown}, is a result of accidental cancellation between the Cabibbo-enhanced semileptonic contribution and the nonleptonic weak-exchange contribution at this order, but this cancellation does not, in general, extend to other mass schemes. However, note that in different mass schemes the relative sizes of the LO and NLO contributions are effectively rearranged leaving the final predictions consistent within uncertainties.  
Contributions induced by the dimension-seven contributions are also large. For $\Xiccpp$ and $\Omcc$, they come with opposite sign relative to the leading spectator terms. It can be expected that the inclusion of currently-missing NLO corrections, and a refined understanding of the corresponding matrix elements, would be important for solidifying the present theoretical predictions.  Our results for the total decay rates, lifetimes, and lifetime ratios are given in table~\ref{tab:results}, and results for predicted semileptonic decay widths are given in table~\ref{tab:results_SL}.

 \begin{table}[ht]
     \centering
     \begin{tabular}{|c|c|c|} \hline
      Observable    & Kinetic & Pole  \\ \hline
       $\Gamma(\Xiccpp)/{\rm ps^{-1}}$     & $3.1\pm 0.5 ^{+0.9}_{-0.6}$  &$3.2\pm 0.4 ^{+0.9}_{-0.7}$  \\
     $\Gamma(\Xiccp)/{\rm ps^{-1}}$     & $17.3\pm 3.4 ^{+4.4}_{-3.6}$ & $16.9\pm 3.3 ^{+4.2}_{-3.5}$\\
     $\Gamma(\Omcc)/{\rm ps^{-1}}$     & $6.7\pm 1.8 ^{+1.2}_{-1.1}$ & $6.1\pm 1.7 ^{+1.3}_{-1.1}$ \\ 
     			\hhline{|=|=|=|}
     $\tau(\Xiccpp)/10^{-13}s$     & $3.2\pm 0.5 ^{+0.8}_{-0.7}$  &$3.2\pm 0.4 ^{+0.9}_{-0.7}$  \\
     $\tau(\Xiccp)/10^{-13}s$     & $0.6\pm 0.1 ^{+0.2}_{-0.1}$ & $0.6\pm 0.1 ^{+0.2}_{-0.1}$\\
     $\tau(\Omcc)/10^{-13}s$     & $1.5\pm 0.4 ^{+0.3}_{-0.2}$ & $1.6\pm 0.5 ^{+0.4}_{-0.3}$ \\ 
     			\hhline{|=|=|=|}
     $\tau(\Xiccp)/\tau(\Xiccpp)$ &$0.22\pm 0.05 ^{+0.04}_{-0.04}$ & $0.22\pm 0.05 ^{+0.04}_{-0.03}$\\
     $\tau(\Omcc)/\tau(\Xiccpp)$ &$0.52\pm 0.13 ^{+0.03}_{-0.02}$ &$0.57\pm 0.15 ^{+0.04}_{-0.03}$ \\ \hline
     \end{tabular}
     \caption{Values for the total decay rates, lifetimes, and the lifetime ratios, in the pole- and kinetic mass schemes. Uncertainties arise from  parametric (first) and scale $\mu$ variations (second). The predictions for the ratios are obtained using the experimental value of $\tau(\Xiccpp)$ for the normalisation according to the formula \eqref{Eq: RatioDef}.}
     \label{tab:results}
 \end{table}

 \begin{table}[ht]
     \centering
     \begin{tabular}{|c|c|c|} \hline
      Observable  & Kinetic & Pole \\ \hline
     $\Gamma_{\rm SL}(\Xiccpp)/{\rm ps^{-1}}$     &$0.84\pm 0.03 ^{+0.05}_{-0.07}$  & $0.70\pm 0.03 ^{+0.07}_{-0.09}$ \\
     $\Gamma_{\rm SL}(\Xiccp)/{\rm ps^{-1}}$     &$0.86\pm 0.03 ^{+0.05}_{-0.07}$  & $0.71\pm 0.04 ^{+0.07}_{-0.09}$\\
     $\Gamma_{\rm SL}(\Omcc)/{\rm ps^{-1}}$     & $1.42\pm 0.56 ^{+0.01}_{-0.09}$ & $1.07\pm 0.54 ^{+0.01}_{-0.10}$ \\ \hline
     \end{tabular}
     \caption{Semileptonic decay widths including both the electron and the muon contributions, in units $\rm ps^{-1}$, in the pole- and kinetic mass schemes. Uncertainties arise from the hadronic (first) and scale variations (second).
     }
     \label{tab:results_SL}
 \end{table}

To date, the only observable to have been measured is the lifetime of $\Xiccpp$ \cite{LHCbXiccpp2018}, with the value given in \eqref{eq:LHCbtXiccpp}.
Our  result, in table~\ref{tab:results}, is consistent with this value for both mass schemes considered in this paper, within uncertainties. As the remaining observables are yet to be measured, we provide them as predictions, to be compared against analysis of LHCb data as and when it becomes available.\footnote{We thank He Jibo for his comments on the ongoing analysis at the LHCb.}

We also report our predictions for the inclusive semileptonic decay widths,
\begin{equation}
    \Gamma_{\rm SL}(\Barycc)\equiv \Gamma(\Barycc \to X e \nu)+\Gamma(\Barycc \to X \mu \nu)\,,
\end{equation}
with the corresponding numerical values given in table~\ref{tab:results_SL}. Note that the semileptonic width $ \Gamma_{\rm SL}(\Omcc)$ turns out to be particularly sensitive to the choice of the charm quark mass scheme. In addition, the hadronic uncertainties for this observable are large, which can be attributed to cancellation between dimension-six and -seven contributions, both of which are Cabibbo-enhanced.\footnote{A similar feature has been observed for the singly charmed $\Omega_c$ baryon \cite{Cheng:2023voj, GMN2022}.} It can be expected that the inclusion of NLO corrections to the dimension-seven Wilson coefficients would improve the numerical stability of this prediction, both by reducing the sensitivity to the choice of charm quark mass scheme and by potentially reducing the overall uncertainties. Moreover, we stress that an experimental measurement of the semileptonic decay widths would provide important insights into the structure of the corresponding hadronic matrix elements.

We now briefly comment on how our results compare to previous predictions. The earliest estimate for the lifetimes dates to the 1980s \cite{Bjorken:1986kfa}, with the expected hierarchy \eqref{eq:ourhierarchy} first noted in \cite{FR1989}. The first comprehensive evaluation of the lifetimes came around a decade later \cite{KLO1998,Likhoded:1999yv,Melic99cc,KL2001,CLLW07cc}, with the most recent studies in \cite{KR2014,KR2018,ChengShi18cc,BLL2018,LL2018}. A complete comparison, including the intermediate analytic expressions, is somewhat challenging, as earlier authors have not always been clear about their conventions, and there are some apparent typographical/sign errors, in e.g.\ eq.~(22) of \cite{BKLO1998}.\footnote{Note that, contrary to the claim in \cite{ChengShi18cc}, there is no discrepancy between their analytic results and those of \cite{Melic99cc}, once corrections in the erratum of the latter paper are taken into account (see also table 1 in \cite{Melic99cc2}). Likewise, the results for lifetimes from \cite{Melic99cc} are incorrectly quoted in table XII of \cite{Cheng:2021qpd}.} One particular source of confusion is how to handle the presence of an additional charm quark; for example, \cite{Melic99cc} introduces the resulting factor of two at the level of the decay width. In our presentation, discussed at length in section~\ref{sec:matrixelements}, such a factor always arises from the evaluation of matrix elements. 
As compared with earlier studies, we have included additional contributions in the $1/m_c$ and $\als$ expansions, in particular the nonleptonic Darwin contribution,  which was not available until recently, and sizeable NLO $\als$ corrections to the dimension-six contributions. 
In addition, most early papers do not provide a comprehensive uncertainty analysis, and can therefore be sensitive to changes in numerical input. 
Nevertheless, all previous studies are broadly consistent with the hierarchy
$\tau(\Xiccp) < \tau(\Omcc) < \tau(\Xiccpp) \,,$ with which we also agree, as shown in eq.~\eqref{eq:ourhierarchy}. 

\begin{figure}
     \centering
     \includegraphics{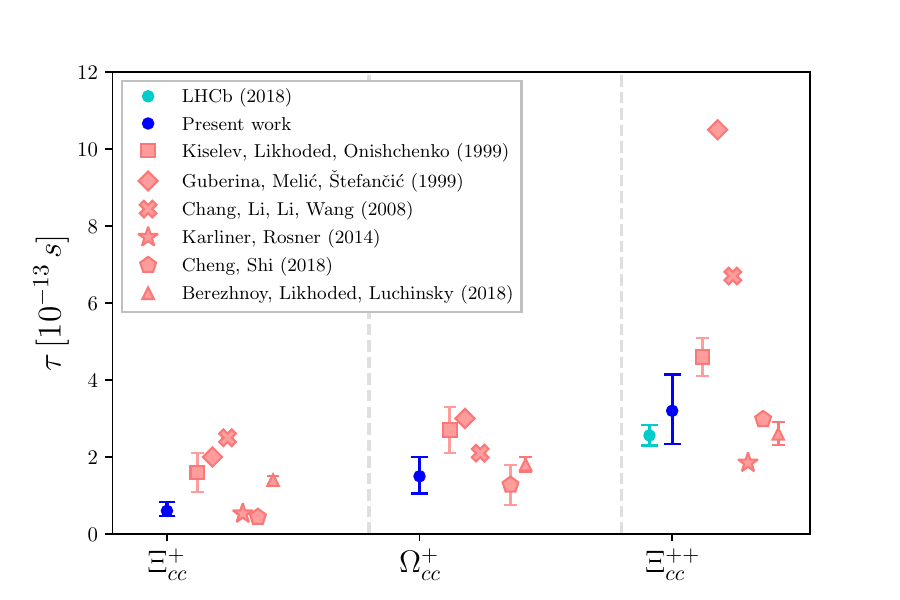}
 \caption{Comparison of our predictions for the lifetimes of $\Xiccp,\, \Omcc,\, \Xiccpp$, in dark blue, with the LHCb measurement of $\tau(\Xiccpp)$ \cite{LHCbXiccpp2018} in cyan, and other theoretical predictions \cite{KLO1998,Likhoded:1999yv, Melic99cc, Melic99cc2,KL2001,CLLW07cc, KR2014, ChengShi18cc, BLL2018} in light red, in chronological order from left to right. Where available, uncertainties in the previous theoretical predictions are also indicated. 
}
     \label{fig:resultslifetimes}
 \end{figure}

In figure~\ref{fig:resultslifetimes}, we summarise our final results for lifetimes, and compare them to previous predictions and the available experimental measurement.

\section{Summary}
\label{sec:conclusion}

In this work, we have provided updates to the predictions for lifetimes of doubly charmed baryons, with newly-available contributions to the decay width included. We also provide, for the first time, predictions for the lifetime ratios, which benefit from reduced uncertainties. Our result for the $\Xiccpp$ lifetime is compatible with the available experimental measurement, while the remaining results presented in this paper serve as predictions for anticipated future results, which can be expected from the analysis of Run 3 LHCb data.

As compared with previous predictions, we have extended the analysis in the following ways:
\begin{itemize}
    \item inclusion of the Darwin contribution in the full decay width, not available for any previous studies;
    \item inclusion of available NLO corrections to leading two- and four-quark contributions;
    \item inclusion of $1/m_c$ corrections to four-quark operators, not available in \cite{KLO1998,BKLO1998,KL2001,Melic99cc,LL2018}, with a fresh consideration of their parametrisation as opposed to \cite{ChengShi18cc} in light of the discussions in \cite{LenzNote:2021,GMN2022};
    \item a more complete analysis of uncertainties, taking hadronic and scale uncertainties into account.
\end{itemize}
Our results provide further support for the hierarchy $\tau(\Xiccp) < \tau(\Omcc) < \tau(\Xiccpp)$, 
which can be tested concretely in future and ongoing collider experiments. 

As with the previous studies of charmed hadrons \cite{LenzNote:2021,GMN2022}, these predictions can evidently be extended in multiple ways: firstly, by adding additional contributions in the $1/m_c$ expansion, which can be expected to converge only slowly; secondly, by consideration of higher-order QCD corrections, which will also serve to reduce the sizeable uncertainty arising from renormalisation scale variations;\footnote{A step in this direction has been made very recently in the work \cite{MMP2023}, while this paper was in the final stages of preparation, which computed the NLO contribution to the $1/m_c^2$ terms in the full decay width for the first time. This will be particularly useful in the case of singly charmed hadrons, although in the present study, this effect will be dominated by hadronic uncertainties.} and, thirdly, by greater control of the hadronic inputs, which also provide a large uncertainty. 

Most importantly, however, these predictions should serve as motivation for a renewed experimental interest in inclusive doubly charmed baryon decays. As was noted in \cite{LenzNote:2021,GMN2022}, there are some tensions between the theoretical predictions and experimental results for singly charmed hadron decays; while the overall pattern is one of qualitative agreement, it is clear that doubly charmed baryon observables provide a further test of the applicability of the heavy quark expansion to charm decays.

\subsection*{Acknowledgments}
We wish to thank He Jibo for useful comments on experimental searches for doubly charmed baryons at the LHCb. JG wishes to thank Benjam\'{i}n Grinstein for useful discussions at the Zadar 2022 Workshop   ``Quirks in Quark Flavour Physics''. 
Support of the Croatian Science Foundation (HRZZ) project, ``Heavy hadron decays and lifetimes'' IP-2019-04-7094, as well as sponsorship from the Alexander von Humboldt Foundation in the framework of the Research Group Linkage Programme, with funding from the German Federal Ministry of Education and Research, is gratefully acknowledged.
\\ 
 \\

\appendix

\section{Conventions and numerical inputs used in the paper}

\subsection{Conventions}
\label{app:conventions}

Here we gather conventions used in deriving the presentation of equations earlier.
We define the covariant derivative
\begin{align}
    D_{\mu} = \partial_{\mu} - i  g_s A^a_\mu t^a 
\end{align}
and thus
\begin{align}
    \left[ i D_\mu, i D_\nu \right] = i g_s G_{\mu \nu} \,.
    \end{align}
The equation of motion for the gluon field strength is therefore
\begin{align}
    \left[ D_\mu, G^{\mu \nu} \right] = - g_s t^a \bar{q}_f t_a \ga^\nu q_f \,, 
    \label{eq:EOM}
\end{align}
with an implicit sum over the colour index $a$ and all the flavours $f$. Our convention for the chromoelectric and chromomagnetic fields is given as  
\begin{equation}
    G^{i0} = E^i,\qquad\qquad B^i = -\frac{1}{2} \eps_{ijk}G^{jk}\,,
\end{equation}
with the convention for the totally  antisymmetric symbol $\epsilon_{123}=+1$.
In the Darwin term, the covariant derivative $\vec{{\cal D}}$ in the adjoint representation acts on the chromoelectric field as 
\begin{align}
( \vec{{\cal D}} \cdot \vec{E} ) = ( \vec{\partial} t^a - g f^{abc} t^b \vec{A}^c ) \cdot\vec{E}^a
\end{align}

\subsection{Numerical inputs}
\label{app:inputs}
In this section, we collect the numerical inputs used in this work. Values of the inputs are taken from PDG \cite{PDG2020,PDG2022} except where stated. Parameters also used in our previous paper \cite{GMN2022} have the same value. 

\begin{table}[ht]
	\centering
	\begin{tabular}{ |c|c| }
		\hline
		Parameter & Numerical value \\
			\hhline{|=|=|}
		$\overline{m}_s(2 \GeV)$ & $0.093\GeV$ \\
		\hline
		$\overline{m}_c(\overline{m}_c)$ \cite{FLAG2019,FLAG2021} & $1.280(13)\GeV$  \\
		\hline
		$\overline{m}_b(\overline{m}_b)$ \cite{FLAG2019,FLAG2021} & $4.198(12)\GeV$  \\
		\hline
		$\alpha_s{(m_Z)}$ & $0.1180(7)$ \\
			\hline
		$m_Z$ & $91.1876\GeV$ \\
			\hline
		$m_{\mu}$ & $0.105658\GeV$ \\
		\hline
	\end{tabular}
	\caption{\small Values of input parameters used in the numerical analysis. Uncertainties in the final digit(s), which are neglected in this study, are given in brackets. 
	The values of $\overline{m}_c(\overline{m}_c)$ and $\overline{m}_b(\overline{m}_b)$ are the Flavour Lattice Averaging Group (FLAG) \cite{FLAG2019,FLAG2021} averages.
 }
	\label{tab:inputs}
\end{table}

\begin{table}[ht]
    \centering
    \begin{tabular}{|c|c|c||c|c|}
    \hline
         & $\Xiccpp,\,\Xiccp$ & $\Omcc$ & $\Xiccppstar,\,\Xiccpstar$ & $\Omccstar$ \\
    \hline 
      masses from \cite{BDMO2014}   & 3.610(32) & 3.738(28) & 3.692(35) & 3.822(30)  \\
      Experimental masses  &  3.62155(38) \cite{LHCbXiccpp2019} & N/A  &  N/A & N/A \\
    \hline 
    \end{tabular}
    \caption{Masses, in $\GeV$, of doubly charmed baryons, and their resonances, taken from the lattice study in \cite{BDMO2014}, with statistical and systematic uncertainties added in quadrature. Only the $\Xiccpp$ has been confirmed experimentally. Given the expectation that isospin splitting should not lead to large mass differences, e.g.\ \cite{KR2017}, the reported mass of $\Xiccp$ in \cite{SELEXXiccp2002,SELEXXiccp2004} is not used in this study. }
    \label{tab:Bcc_masses}
\end{table}

\begin{table}[ht]
    \centering
    \begin{tabular}{|c|c|c|c|} \hline
         & $D^\pm$ & $D^0$ & $D_s$ \\ \hline
    $m_M$     & $1.86966(5)$ & $ 1.86484(5) $ & $ 1.96835(7)$ \\
    $f_M$     & $0.2120(7)$ & $0.2120(7)$ & $0.2499(5) $ \\\hline
    \end{tabular}
    \caption{\small Masses and decay constants of $D$ mesons in GeV, from the latest PDG \cite{PDG2020} and FLAG \cite{FLAG2019,FLAG2021} values. Uncertainties in the final digit(s), which are neglected in this study as they are dominated by other effects, are given in brackets.}
    \label{tab:Dmes_masses}
\end{table}

All quark masses in the matrix elements of four-quark operators are taken to be their constituent masses in baryons and mesons, which take different values and are obtained from fits to experimentally determined hadron masses \cite{GR1981,KR2014}. Specifically, the constituent quark masses in mesons are
\begin{equation}
m_{u,d}^\Mes = 310 \MeV\,, \quad m_s^\Mes = 483 \MeV \,, \quad m_c^\Mes = 1663.3 \MeV \,, \quad m_b^\Mes = 5003.8 \MeV \,,  
\label{eq:mqeffmes}
\end{equation}
and in baryons, where the expected quark masses are somewhat larger, are
\begin{equation}
m_{u,d}^\Bary = 363 \MeV\,, \quad m_s^\Bary = 538 \MeV\,, \quad m_c^\Bary = 1710.5 \MeV \,, \quad m_b^\Bary = 5043.5 \MeV \,.
\label{eq:mqeffbary}
\end{equation}

\section{Contributions to decay width from four-quark operators}
\label{app:4q}
Here we compile analytic expressions for the leading-order spectator contributions to the inclusive decay width \eqref{eq:HQEsystematic}, as given by eqs.~(\ref{eq:Xiccppspectator}-\ref{eq:Omccspectator}). They are (e.g.\ \cite{LenzRauh2013,GMN2022})
\begin{align}
\widehat{\Ga}^{q}_{6,\intp}(x_1, x_2)&=\frac{\Ga_0}{2m_M}\frac{16\pi^2\sqrt{\la}}{m_Q^3}\Bigg\{\Big[\big((x_1-x_2)^2+x_1+x_2-2\big)(2C_1 C_2+\NC C_2^2)\Big]O^q_1\nonumber \\
&-\Big[ 2\big(2(x_1-x_2)^2-x_1-x_2-1\big)(2C_1 C_2+\NC C_2^2)\Big]O^q_2\nonumber \\
&+\Big[\big((x_1-x_2)^2+x_1+x_2-2\big)C_1^2\Big] \tilde{O}^q_1-2\Big[\big(2(x_1-x_2)^2-x_1-x_2-1\big)C_1^2\Big]\,\tilde{O}^q_2\Bigg\}\,, \nonumber  \pagebreak[2]\\
\widehat{\Ga}^q_{6,\exc}(x_1,x_2)&=\frac{\Gamma_0}{2m_B}\frac{16\pi^2}{m_Q^3} \left(2 \NC \sqrt{\la} (1-x_1-x_2) \right)\Big\{\Big[2 C_1 C_2\Big]O^q_1+\Big[C_1^2+C_2^2\Big]\tilde{O}^q_1\Big\}\,,\nonumber \pagebreak[2] \\
\nonumber\\
\widehat{\Ga}^{q}_{6,\intm}(x_1,x_2)&=\widehat{\Ga}^{q}_{6,\intp}(x_1, x_2)\big|_{C_1\longleftrightarrow C_2}\,,\nonumber \\
\nonumber\\
\widehat{\Ga}^{q,\text{SL}}_{6,\intp}(x_\ell,0)&=\widehat{\Ga}^{q}_{6,\intp}(x_\ell,0)\big|_{C_1 \to 0, C_2 \to 1, N_C \to 1}\,,
\label{eq:explicitDim6Bar}
\end{align}
and for dimension seven,
\begin{align}
    \widehat{\Ga}^{q}_{7,\intp}&=\frac{\Ga_0}{2m_M}\frac{16\pi^2\sqrt{\la}}{m_Q^4}(2 C_1 C_2 + N_C C_2^2) \bigg\{2 \left[2(x_1- x_2)^2  - x_1 -x_2 -1\right] \left(P_1^q +{P_1^{q}}^{\dagger}\right) \nonumber \\
    &{}+\frac{2}{\lambda}\bigg[(x_1+x_2-1)\Big((x_1-x_2)^2+x_1+x_2-2\Big)+\lambda\Big(2(x_1-x_2)^2+x_1+x_2\Big)\bigg]\OpsevenP{2}{q}\nonumber \\
    &{}+\frac{4}{\lambda}\bigg[(1-x_1-x_2)\Big(\lambda+(x_1-x_2)^2+x_1+x_2-2\Big) \nonumber \\ & \qquad {} +\lambda\Big(1+2x_1+2x_2-6(x_1-x_2)^2\Big)\bigg]\OpsevenP{3}{q}\bigg\} \nonumber \\
     &{} + \bigg\{\OpsevenP{i}{q} \to \OpsevenPt{i}{q}, (2 C_1 C_2 + N_C C_2^2) \to C_1^2 \bigg\} \,, \nonumber \\
    \widehat{\Ga}^q_{7,\exc}&=\frac{\Gamma_0}{2m_B}\frac{16\pi^2}{m_Q^4} \Big[\frac{12\left((1-x_1-x_2)^2+(x_1+x_2)\la\right)}{\sqrt{\la}}\Big]\Big\{\Big[2 C_1 C_2\Big]P^q_2+\Big[C_1^2+C_2^2\Big]\tilde{P}^q_2\Big\}\,, \nonumber \\
    \nonumber\\
    \widehat{\Ga}^{q}_{7,\intm}&=\widehat{\Ga}^{q}_{7,\intp}\big|_{C_1\longleftrightarrow C_2}\,, \nonumber \\
    \nonumber\\
    \widehat{\Ga}^{q,\text{SL}}_{7,\intp}(x_\ell,0)&=\widehat{\Ga}^{q}_{7,\intp}(x_\ell,0)\big|_{C_1 \to 0, C_2 \to 1, N_C \to 1}\,,
\label{eq:explicitDim7Bar}
\end{align}
with the $(x_1,x_2)$ dependence suppressed. In both eqs.~\eqref{eq:explicitDim6Bar} and \eqref{eq:explicitDim7Bar}, $\la \equiv \la(x_1,x_2) = 1 + x_1^2 + x_2^2 - 2(x_1 + x_2 + x_1 x_2) $ is the K\"all\'en function, while $x_i$ denote the dimensionless ratio
\begin{equation}
    x_q = \frac{m_q^2}{m_c^2}\,,\qquad x_\ell = \frac{m_\ell^2}{m_c^2} \,. 
\end{equation}
We treat the $u,\, d,\, e$, and $\nu_i$ as massless, so that only $x_s$ and $x_\mu$ are nonzero.

\bibliographystyle{JHEP.bst}
\bibliography{References_Doubly_Charmed.bib}

\end{document}